\documentclass[prb,twocolumn,showpacs,floatfix]{revtex4}

\usepackage{graphicx}

\begin{document}

\title{
   Interaction and dynamical binding of spin
   waves or excitons in quantum Hall systems}

\author{
   Arkadiusz W\'ojs,$^{1,2}$ Anna G{\l}adysiewicz,$^1$
   Daniel Wodzi\'nski,$^1$ and John J. Quinn$^2$}

\affiliation{
   $^1$Wroclaw University of Technology, 50-370 Wroclaw, Poland\\
   $^2$University of Tennessee, Knoxville, Tennessee 37996, USA}

\begin{abstract}
Interaction between spin waves (or excitons) moving in the 
lowest Landau level is studied using numerical diagonalization.
Becuse of complicated statistics obeyed by these composite 
particles, their effective interaction is completely different 
from the dipole--dipole interaction predicted in the model of 
independent (bosonic) waves.
In particular, spin waves moving in the same direction attract 
one another which leads to their dynamical binding.
The interaction pseudopotentials $V_{\uparrow\uparrow}(k)$ 
and $V_{\uparrow\downarrow}(k)$ for two spin waves with equal 
wavevectors $k$ and moving in the same or opposite directions
have been calculated and shown to obey power laws $V(k)\propto
k^\alpha$ at small $k$.
A high value of $\alpha_{\uparrow\uparrow}\approx4$ explains 
the occurrence of linear bands in the spin excitation spectra 
of quantum Hall droplets.
\end{abstract}
\maketitle

\section{Introduction}

Description of interactions and correlations between excitons
\cite{haug93} (electron-hole pairs, $X=e+h$) is somewhat 
problematic because of their complicated statistics.
Being pairs of fermions, the excitons obey Bose statistics 
under a ``full'' exchange and, consequently, condense into 
a Bose--Einsetin ground state at sufficiently low density.
\cite{keldysh68} 
However, their composite nature comes into play when the 
excitons overlap and ``partial'' exchanges (of only a pair 
of electrons or holes) can occur.
And, unlike for charged complexes (such as trions, $X^-=2e+h$) 
naturally separated by the Coulomb repulsion, the overlaps 
between neutral excitons can often be significant.

In the absence of a magnetic field $B$, exciton correlations 
have been discussed\cite{okamura02} in connection with 
four-wave mixing experiments that involve two-photon 
absorption.\cite{shah93,feuerbacher91,baars98,borri99}
Here, we will consider 2D systems in the high-$B$ limit, 
so-called ``quantum Hall systems.''\cite{prange87} 
While the bosonization scheme for excitons confined to the 
lowest Landau level (LL$_0$) has recently been proposed,
\cite{doretto04} we will concentrate on the numerical 
results for the $X$--$X$ interaction pseudopotential.

In LL$_0$, a well-known statistics/correlation effect is 
the decoupling and condensation of $k=0$ excitons in the 
ground state of interacting electrons and holes.\cite{lerner81}
It can be interpreted in terms of an inter--exciton
($X$--$X$) exchange attraction exactly compensating for 
a decrease in the intra-exciton ($e$--$h$) attraction due 
to the phase space blocking for the coexisting identical 
constituent fermions.

The exciton condensation in LL$_0$ results from the mapping 
of an $e$--$h$ system onto a two--spin system with spin-symmetric
interactions.\cite{macdonald90} 
The ``hidden'' $e$--$h$ symmetry corresponding to the 
conservation of the total spin and responsible for exciton 
condensation holds in LL$_0$ because there the electron and 
hole orbitals are identical despite different effective masses 
(in experimental systems with finite width, this also requires 
symmetric doping to avoid normal electric field that would 
split the $e$ and $h$ layers).

The mapping between $e$--$h$ and two-spin systems makes 
interband excitons in an empty LL$_0$ equivalent to spin 
waves (SW's) in a filled LL$_0$, i.e., in the quantum Hall 
ground state with the filling factor $\nu=1$.
A SW (or spin exciton) consists of a hole in the spin-polarized 
LL$_0$ and a reversed-spin electron in the same LL$_0$.
Although excitons and SW's in LL$_0$ are formally equivalent
and the conclusions of Ref.~\onlinecite{doretto04} and ours 
apply to both complexes, they are relevant for two different 
types of experiments (photoluminescence and spin relaxation).

Being charge-neutral, excitons move along straight lines and 
carry a linear wavevector $k$ even in a magnetic field $B$.
The origin of their (continuous) dispersion\cite{gorkov68} 
$\varepsilon(k)$ in LL$_0$ is not the (constant) $e$ or $h$ 
kinetic energy, but the dependence of an average $e$--$h$ 
separation on $k$.
A moving exciton carries an electric dipole moment $d$, 
proportional and orthogonal to both $k$ and $B$.

For a pair of moving excitons, one could think that the dominant 
contribution to their interaction $V({\mathbf k}_1,{\mathbf k}_2)$ 
would be the dipole--dipole term,\cite{olivares01} specially at 
small values of $k_1$ and $k_2$, when this term is too weak on 
the scale of $\varepsilon(k)$ to cause a significant polarization 
of the $X$ wavefunctions.
Such assumption would lead to the repulsion between excitons 
moving in the same direction.

However, we show that this assumption is completely false 
because of the required (anti)symmetry of the wavefunction 
of overlapping excitons under exchange of individual constituent 
electrons or holes.
This statistics/correlation effect is significant even at small 
$k$, and it reverses the sign of the $X$--$X$ interaction, 
compared to the dipole--dipole term.
Specifically, excitons moving in the same direction attract one 
another, and the ground state of a pair of excitons carrying 
a total wavevector ${\mathbf k}$ is a (dynamically) bound 
state with ${\mathbf k}_1={\mathbf k}_2={1\over2}{\mathbf k}$.

The $X$--$X$ interaction pseudopotential is calculated 
numerically for two special cases: ${\mathbf k}_1=\pm{\mathbf k}_2$, 
corresponding to a pair of excitons moving with equal wavevectors 
$k_1=k_2\equiv k$ in the same ($\uparrow\uparrow$) and opposite 
($\uparrow\downarrow$) direction.
In addition to the sign reversal, we find that the inclusion of 
the statistics effects leads to the significant weakening of the 
$X$--$X$ interaction, specially at small $k$ (e.g., for the 
$\uparrow\uparrow$ configuration. we find a $V\propto k^4$ 
power-law behavior).

The near vanishing of the interaction between excitons moving 
in the same direction explains the occurrence of nearly linear 
multi-exciton bands found numerically in the spin-excitation 
spectra of finite-size quantum Hall droplets
\cite{palacios94,spectral} and of extended quantum Hall systems.
\cite{skyrmion}
And the attractive character of this interaction explains the
slightly convex shape of these bands, which for a confined 
droplet leads to the oscillations of the total spin as a 
function of the magnetic field.\cite{palacios94,spectral}

\section{Model}

\begin{figure}
\resizebox{3.4in}{1.74in}{\includegraphics{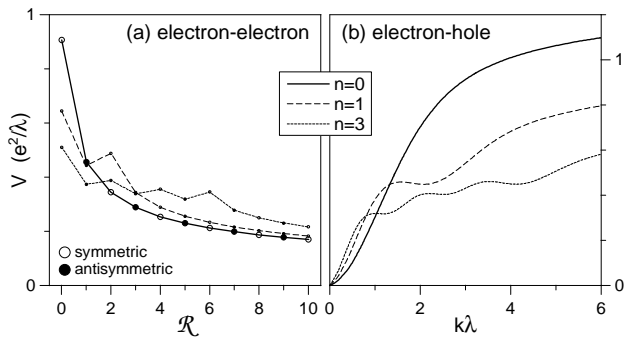}}
\caption{
\label{fig1}
   Electron-electron (a) and electron-hole (b) pseudopotentials 
   in the $n$th ($n=0$, 1, and 3) LL. $V$ is the pair interaction
   energy, $\mathcal{R}$ is the relative pair angular momentum, 
   $k$ is the total pair wavevector, and $\lambda$ is the magnetic 
   length.}
\end{figure}

We consider spin excitations at the filling factor $\nu=1$, i.e., 
in a system of $N$ electrons half-filling the lowest Landau level 
(LL$_0$) single-particle angular momentum ($l$) shell with two-fold 
spin degeneracy and the orbital degeneracy $g\equiv2l+1=N$.
The interaction among the electrons in the Hilbert space restricted 
to LL$_0$ is entirely determined by Haldane pseudopotential
\cite{haldane87} defined as pair interaction energy $V_{ee}$ as 
a function of relative pair angular momentum $\mathcal{R}$ and 
plotted in Fig.~\ref{fig1}(a).
The even and odd values of $\mathcal{R}$ correspond to symmetric 
and antisymmetric pair wavefunction, i.e., to the singlet and 
triplet pair spin state, respectively.
Assuming large cyclotron gap $\hbar\omega_c$ between LL's 
(compared to the Zeeman gap $E_Z$ and the interaction energy 
scale $e^2/\lambda$, where $\lambda=\sqrt{hc/eB}$ is the magnetic 
length), similar low-energy excitations of electrons at larger 
odd integral values of $\nu=2n+1$ occur only in the half-filled 
LL$_n$, and the only difference compared to the $\nu=1$ case is 
a different form of $V(\mathcal{R})$, as shown in 
Fig.~\ref{fig1}(a) for $n=1$ and 3.

The two-spin system of $N=N_\downarrow+N_\uparrow$ electrons can be 
mapped onto that of $K_e=N_\uparrow$ spin-$\uparrow$ electrons and 
$K_h=N-N_\downarrow$ of spin-$\downarrow$ holes.\cite{macdonald90} 
At $\nu=1$, $K_e=K_h\equiv K$.
The electrons and holes obtained through such mapping are both 
spin-polarized, and their (equal) $e$--$e$ and $h$--$h$ interactions 
are determined by the pseudopotential parameters $V_{ee}(\mathcal{R})$
corresponding only to odd values of $\mathcal{R}$.
The effective $e$--$h$ interaction depends on $V_{ee}(\mathcal{R})$ 
at both even and odd values of $\mathcal{R}$, but it can be described 
more directly by an $e$--$h$ pseudopotential (pair $e$--$h$ energy 
$V_{eh}$ as a function of pair wavevector $k$) plotted in 
Fig.~\ref{fig1}(b).
In LL$_0$, both $e$--$e$ and $e$--$h$ pseudopotentials are monotonic,
while in higher LL's they have oscillations reflecting additional
nodes of the single-particle wavefunctions.

Because of the exact mapping between two-spin and two-charge systems,
all results discussed here are in principle applicable to systems of 
conduction electrons and valence holes.
This equivalence is true for ideal systems (with zero layer width $w$ 
and no LL mixing) considered here.
However, in realistic interband systems (realized e.g.\ by optical 
excitation of an electron gas) the $e$ and $h$ wavefunctions are usually 
different both in the plane of motion (because of mass-dependent LL 
mixing) and in the normal direction (because of mass-dependent density 
profiles $\varrho(z)$ and a spatial separation of $e$ and $h$ planes 
induced by an electric field produced by a charged doping layer).
Therefore, the ``hidden symmetry'' is usually broken in experimental 
$e$--$h$ systems, while the equivalent conservation of the total spin 
$S$ is easily realized in the corresponding two-spin systems.

\section{Spin-excitation spectrum at $\nu=1$}

An intriguing feature known to occur in the spin-excitation spectrum 
at $\nu=1$ is the low-energy band that is linear in spin and angular 
momentum.
It was first identified in finite size quantum Hall droplets,
\cite{palacios94} and later discussed\cite{spectral} in Haldane 
spherical geometry,\cite{haldane83} convenient in modeling infinite, 
translationally invariant systems.
\begin{figure}
\resizebox{3.4in}{1.74in}{\includegraphics{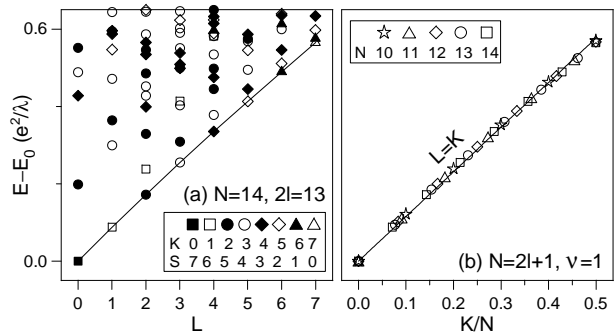}}
\caption{
\label{fig2}
   (a) 
   Energy spectrum (interaction energy $E$ {\sl vs.}\ total 
   angular momentum $L$) of $N=14$ electrons calculated on 
   a sphere for $2l+1=N$ (at filling factor $\nu=1$).
   $S$ is the total spin, $K={1\over2}N-S$, and $\lambda$ 
   is the magnetic length.
   (b)
   Low-energy $L=K$ band for different $N$ as a function of 
   $\zeta=K/N$.}
\end{figure}
As shown in Fig.~\ref{fig2}(a) obtained for $N=14$ electrons on a sphere,
the lowest state at each total angular momentum $L$ has the total 
spin $S$ corresponding to $K={1\over2}N-S$ (the number of spin flips 
relative to the polarized ground state) equal to $L$.
This band is nearly linear in $L$ and thus it can be interpreted
as containing states of $K$ ordered and noninteracting SW's, each 
carrying angular momentum $\ell=1$ and energy $\varepsilon_\ell=
V_{eh}(k_\ell)$, where $k_\ell=\ell/R$ (and $R$ is the sphere radius).
Ordering means here that the angular momentum vectors of the $K$ SW's 
are all parallel to give a total $L=K\ell$, i.e., that all SW's move 
in the same direction along the same great circle of the sphere.
On a plane (corresponding to $R\rightarrow\infty$), this corresponds 
to $K$ SW's moving in parallel along a straight line, each with an 
infinitesimal wavevector $k_\ell$.

Scaling of this $L=K$ band with the size of the system is shown in 
Fig.~\ref{fig2}(b), where we overlay the data for different $N\le14$.
The excitation energy $E$ appears be a (nearly size-independent) 
linear function of ``spin polarization'' $\zeta=K/N$.
Assuming exact decoupling of SW's in this band, $E(\zeta)\equiv 
K\varepsilon_\ell$ can be extrapolated to the planar geometry, 
where the SW dispersion is\cite{gorkov68}
\begin{equation}
   V_{eh}(k)=\sqrt{\pi\over2}\left(1-e^{-\kappa^2}I_0(\kappa^2)
   \right){e^2\over\lambda},
\end{equation}
with $\kappa={1\over2}k\lambda$ and $I_0$ being the modified Bessel 
function of the first kind.
For small $k_\ell$, 
\begin{equation}
   \varepsilon_\ell\equiv V_{eh}(k_\ell)\approx
   \sqrt{\pi\over2}\,\kappa_\ell^2\,{e^2\over\lambda}.
\end{equation}
Substituting $k_\ell\lambda=\ell/R$, $R=\sqrt{Q}\lambda$ (where 
$2Q$ is the magnetic monopole strength; $2Q\cdot hc/e=4\pi R^2B$), 
$l=Q$ for the lowest electron shell (LL), and, at $\nu=1$, $N=g
\equiv2l+1$, we have $k_\ell\lambda=\sqrt{2/N}$, and finally
\begin{equation}
   E(\zeta)=\zeta\sqrt{\pi\over8}\,{e^2\over\lambda}.
\end{equation}
This slope is much smaller from the one in Fig.~\ref{fig2}(b) due 
to finite-size/curvature errors on a sphere, particularly significant 
at small $k_\ell$.
The total wavevector $k=L/R=Kk_\ell$ for the $L=K$ band scales as
\begin{equation}
\label{eqdivk}
   k\lambda=\sqrt{2N}\zeta,
\end{equation}
i.e., on a plane is it divergent.
Therefore, $E(\zeta)$ is a lower bound for the actual excitations 
at a given $\zeta$ that will have large but finite $k$.

\section{Effective SW--SW interaction}

\begin{figure}
\resizebox{3.4in}{1.74in}{\includegraphics{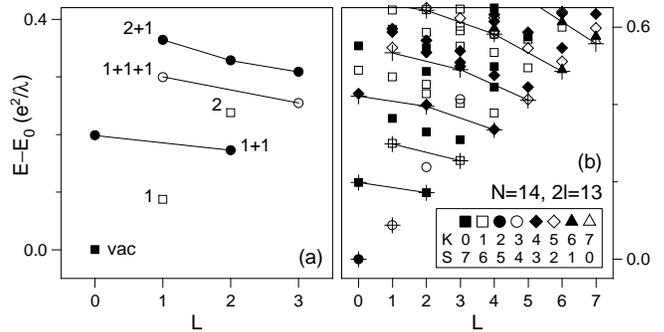}}
\caption{
\label{fig3}
   (a) 
   Low-energy part of Fig.~\ref{fig2}(a).
   Labels indicate angular momenta $\ell$ of the 
   (interacting) SW's in each 14-electron state.
   (b)
   Approximate energies (``$+$'') of 14-electron states 
   containing a number of interacting SW's each with 
   $\ell=1$ compared with the exact spectrum of 
   Fig.~\ref{fig2}(a).}
\end{figure}

Regardless of divergence of $k$ in Eq.~(\ref{eqdivk}), the (nearly) 
linear behavior of $E(K)$ suggests decoupling of SW's in the $L=K$
band and invokes a more general question of interaction between
SW's in the lowest (or higher) LL's.
Unlike their number $K={1\over2}N-S$, the individual angular momenta 
of interacting SW's are not conserved.
For example, a pair of SW's both with $\ell=1$ and with the total 
angular momentum $L=2$ are coupled to a pair with the same $L$ but 
with different $\ell=1$ and 2; these two configurations being denoted 
as $\left|1+1;2\right>$ and $\left|1+2;2\right>$.
However, unless the single-SW energies $\mathcal{E}$ of such coupled 
configurations (here, $\mathcal{E}=2\varepsilon_1$ and $\varepsilon_1
+\varepsilon_2$) are close, this coupling can be effectively 
incorporated into the SW--SW interaction.
In Fig.~\ref{fig3}(a) we have made such assignment for the lowest 
excitations of the 14-electron spectrum.
Following this assignment, we can extract not only the (exact) 
single-SW energies, $\varepsilon_L=E[L]-E_0$, but also the parameters 
of an effective SW--SW interaction pseudopotential, $V[\ell+\ell';L]=
E[\ell+\ell';L]-\varepsilon_\ell-\varepsilon_{\ell'}-E_0$.
Using these two-SW interaction parameters one can describe 
interactions in the states of more than two SW's.

Let us demonstrate it on a simple example of $K$ SW's each with 
$\ell=1$.
In this case, there are only two pair-SW states, at $L=0$ and 2, 
corresponding to the relative (with respect to the center of mass
of the two SW's) angular momenta $\mathcal{R}\equiv2\ell-L=2$ and 0
(SW's are pairs of fermions, and thus for two SW's with equal $\ell$,
${\mathcal R}$ must be even as for two identical bosons).
Thus, there are only two interaction parameters, in a 14-electron 
system equal to $V_2\equiv V[1+1;0]=0.0236\,e^2/\lambda$ and 
$V_0\equiv V[1+1;2]=-0.0026\,e^2/\lambda$ (note that for the 
subscripts in $V_0$ and $V_2$ we use notation $V_\mathcal{R}$ 
and not $V_L$).

The total energy of the state $\Psi$ of $K$ SW's, $E=E_0+K
\varepsilon_\ell+U$, contains the inter-SW interaction energy 
that can be expressed as
\begin{equation}
\label{eqv1}
   U={K\choose2}\sum_\mathcal{R}\mathcal{G}_\mathcal{R}V_\mathcal{R}.
\end{equation}
Here, $\mathcal{G}_\mathcal{R}$ are the pair amplitudes
\cite{haldane87,parentage} (pair-correlation functions) that measure 
the number of SW pairs with a given $\mathcal{R}$ (for brevity, we 
omit index $\Psi$ in $E$, $U$, and $\mathcal{G}_\mathcal{R}$).
They are normalized, $\sum_\mathcal{R}\mathcal{G}_\mathcal{R}=1$, 
and satisfy an additional sum rule that on a sphere has the form
\cite{sum-rule}
\begin{equation}
\label{eqsr}
   L(L+1)+K(K-2)\,\ell(\ell+1)={K\choose2}
   \sum_\mathcal{R} \mathcal{G}_\mathcal{R}\,\mathcal{L}
   (\mathcal{L}+1),
\end{equation}
where $L$ and $\mathcal{L}\equiv 2\ell-\mathcal{R}$ are the total 
and pair SW angular momenta, respectively.

For $\ell=1$, there are only two pair amplitudes, $\mathcal{G}_0$
and $\mathcal{G}_2$, and hence they are independent of the SW--SW 
interaction and can be completely determined from Eq.~(\ref{eqsr}).
This allows expression of $\mathcal{G}_\mathcal{R}$ and, using the 
values of $V_\mathcal{R}$ and Eq.~(\ref{eqv1}), of $U$ and $E$ 
as a function of $K$ and $L$,
\begin{eqnarray}
\label{eqv2}
   U&=&{L(L+1)+2K(K-2)\over6}(V_0-V_2)\nonumber\\
    &+&{K(K-1)\over2}V_2.
\end{eqnarray}
For $L=K$ this gives $\mathcal{G}_2=0$ and $U={1\over2}K(K-1)V_0$,
i.e., the linearity of $E(K)$ depends on the vanishing of $V_0$.
Energies $E(K,L)$ obtained from Eq.~(\ref{eqv2}) for all 
combinations of $L$ and $K$ are compared with the exact 14-electron 
energies in Fig.~\ref{fig3}(b).
Good agreement, especially for the $L=K$ band, justifies 
interpretation of the actual spin excitations in terms of $K$ SW's 
with well-defined $\ell$, interacting through the effective 
SW--SW pseudopotentials.

\section{SW--SW pseudopotential}

This brings up the question of why are the SW's in the $L=K$ band
(nearly) noninteracting (i.e., why is $V_0$ so small compared to 
$V_2$ or $\varepsilon_1$).
And a more general one, what is the pseudopotential describing 
interaction between the SW's.
The SW--SW pseudopotential $V$ depends on the pair of wavevectors, 
${\mathbf k}$ and ${\mathbf k}'$.
However, in extension of $V_0$ and $V_2$ in Eq.~(\ref{eqv2}), we 
will only consider two special cases: $V_{\uparrow\uparrow}(k)$ 
and $V_{\uparrow\downarrow}(k)$, corresponding to two SW's with 
equal wavevectors $k$ moving in the same and opposite directions, 
respectively.

\subsection{Independent SW's}

\begin{figure}
\resizebox{3.4in}{1.74in}{\includegraphics{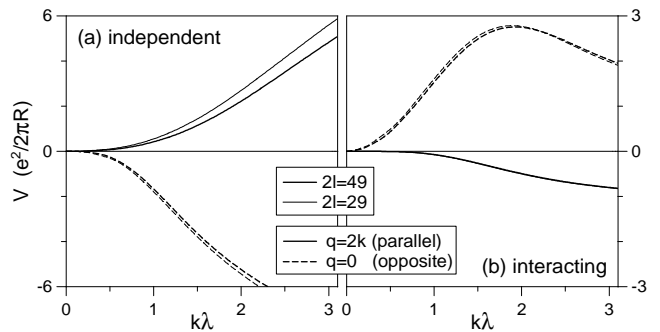}}
\caption{
\label{fig4}
   SW--SW pseudopotentials (two-SW interaction energy $V$ 
   {\sl vs.}\ single-SW wavevector $k$) for the independent 
   (a) and correlated (b) SW's moving in the same or opposite 
   direction (total pair wavevector $q=2k$ or $0$, respectively).
   $\lambda$ is the magnetic length and $R$ is the sphere radius.}
\end{figure}

A moving SW carries\cite{gorkov68} an in-plane dipole electric 
moment ${\mathbf d}$, with magnitude $d$ proportional to $k$ and 
oriented orthogonally to the direction of ${\mathbf k}$.
For a pair of uncorrelated SW's this implies simple dipole--dipole 
interaction, repulsive for the $\uparrow\uparrow$ configuration,
and attractive for $\uparrow\downarrow$.
Indeed, in Fig.~\ref{fig4}(a) we plot $V_{\uparrow\uparrow}(k)$ 
and $V_{\uparrow\downarrow}(k)$ showing such behavior.
Moreover, at small $k$ we find a very regular power-law dependence,
\begin{equation}
\label{eqpwr1}
   V_{\uparrow\uparrow}(k)\sim 
   0.42\,(k\lambda)^{5\over2}{e^2\over2\pi R}.
\end{equation}
The curves in Fig.~\ref{fig4}(a) have been calculated as an 
expectation value of the Coulomb interaction in a trial state
$\left|k,k;q\right>$ describing two uncorrelated (independent) 
SW's, each with the wavevector $k$ and with the total wavevector
$q=2k$ ($\uparrow\uparrow$) and $q=0$ ($\uparrow\downarrow$).
Such trial states have been constructed on a sphere in the basis 
of two electrons and two holes in a lowest LL with $l=Q$.
The two electrons (and two holes) are distinguished by different
isospins $\sigma=\pm{1\over2}$. 
A pairing hamiltonian $H_\ell$ is introduced with the $e$--$h$ 
pseudopotential in the form
\begin{equation}
   V_{eh}^{(\ell)}(\sigma_e,\sigma_h,\ell')
   =-\delta_{\sigma_e\sigma_h}\delta_{\ell\ell'}
\end{equation}
and the $e$--$e$ and $h$--$h$ interactions set to zero.
At each total angular momentum $L$, there is exactly one eigenstate 
of $H_\ell$ corresponding to the eigenvalue $-2$.
It describes two independent $e$--$h$ pairs (i.e., excitons or SW's),
one with $\sigma_e=\sigma_h={1\over2}$ and one with $\sigma_e=\sigma_h
=-{1\over2}$, each in an eigenstate of pair angular momentum $\ell$ 
corresponding to the pair wavevector $k_\ell=\ell/R$ (on a sphere, 
describing motion of a charge-neutral pair along a great circle).
The total angular momentum $L$ of two pairs can also be converted
into the total wavevector, $q=L/R$.
We have concentrated on the trial states with $L=2\ell$ and $0$ 
(i.e., with $q=2k_\ell$ and 0), denoted as $\left|k_\ell,k_\ell;
2k_\ell\right>$ and $\left|k_\ell,k_\ell;0\right>$.
They describe two pairs each with the same $k_\ell$ and moving in 
the same and opposite directions, respectively.
Discrete SW--SW pseudopotentials $V_{\uparrow\uparrow}(k_\ell)$ and 
$V_{\uparrow\downarrow}(k_\ell)$ on a sphere have been calculated as 
the expectation value of the inter-SW Coulomb interaction (i.e., the 
total Coulomb energy of the $2e+2h$ state minus the intra-SW $e$--$h$ 
attraction $2\varepsilon_\ell$).
When the sphere curvature $R/\lambda=Q^2$ decreases, the discrete values 
quickly converge to the continuous curves $V_{\uparrow\uparrow}(k)$ and 
$V_{\uparrow\downarrow}(k)$ appropriate for a planar system.
The interpolated curves for the LL degeneracy $2l+1\equiv2Q+1=30$ and 
50 are compared in Fig.~\ref{fig4}(a).
Note that $V$ is plotted as a function of $e^2/2\pi R$ (rather than 
$e^2/\lambda$) what reflects the fact that SW's are extended objects
confined to a great circle of length $2\pi R$ (in contrast to electrons
or holes that are confined to cyclotron orbits of radius $\sim\lambda$).

\subsection{Coupled SW's}

The SW--SW pseudopotentials obtained above describe interaction
between independent SW's (distinguished by isospins $\sigma_e$ and 
$\sigma_h$).
However, the following two correlation effects must be incorporated into 
the effective SW--SW interaction to describe the actual spin excitations 
at $\nu\sim1$ (i.e., the interacting $e$--$h$ systems).

First, the Coulomb (charge--charge) interaction between the SW's breaks 
the conservation of $\ell$ and causes relaxation of the individual SW
wavefunctions and their energies $\varepsilon_\ell$.
This perturbation effect mixes the SW states within the energy range 
$\Delta\varepsilon\sim V$, so it becomes negligible when $V$ is small, 
i.e., at small $k$.
In particular, it does not affect the behavior of $V_{\uparrow\uparrow}
(k)$ at small $k$, responsible for the linearity of the $L=K$ band.

Second, strictly speaking, the SW's are not bosons but pairs of 
fermions, and a wavefunction of two SW's must not only be symmetric
under interchange of the entire SW's, but also antisymmetric under
interchange of two constituent electrons or holes.
The trial paired states $\left|k,k;q\right>$ with $H_\ell=-2$ do not 
obey these symmetry requirements, because $H_\ell$ is isospin-asymmetric 
and hence it does not commute with pair $e$ or $h$ isospins, $\Sigma_e$ 
and $\Sigma_h$.
Therefore, the trial eigenstates of $H_\ell=-2$ are different from 
the properly symmetrized eigenstates of $\Sigma_e=\Sigma_h=1$.
This statistics effect is generally weak for spatially separated 
composite particles, but for the SW's moving along the same line 
(or great circle) it is large and cannot be treated perturbatively
(even at small $k$ when the Coulomb SW--SW interaction is negligible).
At each $L$, the exact form of the ground state in the $\Sigma_e
=\Sigma_h=1$ subspace depends on $\ell$ and on the details of the 
actual (Coulomb) hamiltonian, and so does the average value of 
$H_\ell$ (measuring the actual ``degree of pairing'').
However, as a reasonable approximation one can introduce the 
``maximally paired'' states, defined at each $L$ as the lowest-energy 
state of the pairing interaction hamiltonian $V_{eh}^{(\ell)}$ within 
the $\Sigma_e=\Sigma_h=1$ subspace.

The relaxation of the wavefunctions of the overlapping SW's is 
evident from the analysis of the $e$--$e$ and $h$--$h$ pair 
amplitudes $\mathcal{G}(\mathcal{R})$.
For a pair of different particles, such as electrons or holes 
distinguished by isospin $\sigma$ in the trial state $\left|k,k;
q\right>$, $\mathcal{R}$ can be any integer.
Therefore, $\mathcal{G}_{ee}(\mathcal{R})$ and $\mathcal{G}_{eh}
(\mathcal{R})$ calculated for the independent SW's are positive 
at both even and odd $\mathcal{R}$ (in fact, there is no obvious 
correlation whatsoever between the parity of $\mathcal{R}$ and 
the value of $\mathcal{G}_{ee}$ or $\mathcal{G}_{eh}$).
In contrast, for a pair of identical fermions, such as electrons 
or holes in an actual, interacting state of two SW's, 
$\mathcal{G}_{ee}(\mathcal{R})$ and $\mathcal{G}_{eh}(\mathcal{R})$ 
vanish exactly at all even values of $\mathcal{R}$.
The change of pair amplitudes when going from the trial states 
$\left|k,k;q\right>$ to the actual Coulomb ground states is quite 
dramatic, precluding adequacy of the pseudopotentials of 
Fig.~\ref{fig4}(a) for the description of many-SW systems.

Because of the above relaxation effects, interaction between 
the SW's is not purely a two-body interaction, and thus it 
cannot be completely described by a (pair) pseudopotential
$V(k)$.
In other words, a SW--SW pseudopotential taking these effects 
into account is not rigorously defined.
However, as demonstrated in Fig.~\ref{fig2}(b), many-SW spectra
can be reasonably well approximated using an effective 
pseudopotential obtained for only two SW's.

To determine such effective $V_{\uparrow\uparrow}(k)$ and 
$V_{\uparrow\downarrow}(k)$, we calculate the $2e+2h$ Coulomb 
energy spectra similar to the $K\le2$ part of Fig.~\ref{fig3}(a) 
and make analogous assignments for the $K=2$ states.
The lowest state at each even value of $L=2$, 4, \dots\ 
is interpreted as one of two SW's each with $\ell={1\over2}L$ 
and moving in the same direction.
Similarly, consecutive states at $L=0$ contain two SW's each 
with $\ell=1$, 2, \dots\ and moving in opposite directions.
In both cases, $V(\ell)=E-2\varepsilon_\ell-E_0$.
When $\ell$ is converted into $k_\ell=\ell/R$ and $V$ is plotted 
in the units of $e^2/2\pi R$, the discrete pseudopotentials $V
(k_\ell)$ fall on the continuous curves $V_{\uparrow\uparrow}(k)$ 
and $V_{\uparrow\downarrow}(k)$ that very quickly converge to 
ones appropriate for a planar system when the sphere curvature 
$R/\lambda=Q^2$ is decreased.
The interpolated curves for $2l+1\equiv2Q+1=30$ and 50 are compared 
in Fig.~\ref{fig4}(b), showing virtually no size dependence.
Similar curves were obtained for the ``maximally paired'' states
used instead of actual Coulomb eigenstates.

The justification for the above assignment comes from the 
observation of distinct bands in the low-energy $K=2$ spectrum.
The values of $L$ within each band are consistent with the addition 
of angular momenta of two SW's, $|\ell-\ell'|\le L\le\ell+\ell'$ 
(with the additional requirement that $L-2\ell\equiv\mathcal{R}$ 
be even for $\ell=\ell'$).
In the absence of the SW relaxation, these bands would contain the
eigenstates of $\mathcal{E}\equiv\varepsilon_\ell+\varepsilon_{\ell'}$, 
with the intra-band dispersion reflecting interaction of the independent 
SW's with $\ell$ and $\ell'$.
In the actual spectrum, the bands mix, but remain separated, 
making the assignment possible.
The interband mixing and the resulting changes in the energy 
spectrum are precisely the relaxation effects, effectively 
incorporated into $V(k)$.
For $L=0$ ($\uparrow\downarrow$), the mixing is minimal, because 
the contributing ``independent SW'' configurations $\left|\ell,\ell';
L=0\right>$ must all have $\ell=\ell'$, and thus very different 
single-SW energies $\mathcal{E}$.
For $L=2\ell$ ($\uparrow\uparrow$), mixing between configurations 
$\left|\ell+\delta,\ell-\delta;L=2\ell\right>$ with close values of 
$\mathcal{E}$ can occur, having a stronger effect on the effective 
$V_{\uparrow\uparrow}(k)$.

The main two findings about the effective SW--SW pseudopotentials 
shown in Fig.~\ref{fig4}(b) are the following.
First, the statistics effect turns out so strong as to reverse 
the sign of interaction.
In contrast to the prediction of the model of independent SW's 
with dipole--dipole interaction, the SW's moving in the same 
direction decrease their total energy (what can be interpreted 
as attraction), while the SW's moving in opposite direction 
increase their energy (i.e., repel one another).
Second, the magnitude of the $\uparrow\uparrow$ attraction at 
small $k$ is greatly reduced compared to Eq.~(\ref{eqpwr1}).
It can also be approximated by a power-law dispersion, but 
with a much higher exponent and a much smaller prefactor,
\begin{equation}
\label{eqpwr2}
   V_{\uparrow\uparrow}(k)\sim 
   -0.069\,(k\lambda)^4{e^2\over2\pi R}.
\end{equation}
Although the near vanishing of $V_{\uparrow\uparrow}$ at small 
$k$ was anticipated from the linearity of the $L=K$ band in 
Fig.~\ref{fig2}, the negative sign and large exponent are rather 
surprising and of a wider consequence.
It may be worth stressing that the identified attraction between 
$N$ SW's (or interband excitons) moving in the same direction is
too weak to induce a stable bound ground state, with the total 
energy lower than $N$ times ground state energy of a single 
SW/exciton.
Therefore, it does not contradict a well-known fact that the 
ground state of $N$ electrons and $N$ holes in the lowest LL 
is a multiplicative state\cite{lerner81,macdonald90} of $N$ 
SW's/excitons each with $k=0$ (in particular, a biexciton is 
unstable toward breaking up into two $k=0$ excitons, while 
the energy of $N$ SW's is never lower than $N\varepsilon_0=0$, 
and so the $\nu=1$ ground state is spontaneously polarized).
However, for two or more SW's/excitons carrying a conserved 
total wavevector $q>0$, the convex shape of $V_{eh}(k)$ causes 
equal distribution of $q$ among all SW's/excitons, and the 
SW--SW or $X$--$X$ attraction binds them together.
Such a moving multi-SW/exciton can only break up (into 
separate SW's/excitons) through an inelastic collision taking 
away its wavevector.
This dynamical binding will affect spin relaxation (for the 
SW's) or photoluminescence (for the excitons) of an electron 
gas, but the relevant spectra are yet to be calculated.

\section{Conclusion}

We have studied interaction between moving SW's (excitons) in the 
lowest LL.
For a pair of SW's with equal wavevectors $k$ and moving in the same 
($\uparrow\uparrow$) or opposite ($\uparrow\downarrow$) directions, 
the effective interaction pseudopotentials $V_{\uparrow\uparrow}(k)$ 
and $V_{\uparrow\downarrow}(k)$ have been calculated numerically.
They account for relaxation of overlapping SW's due to the Fermi 
statistics of constituent (reversed-spin) electrons and (spin-) 
holes, and differ completely from the prediction for independent 
SW's interacting through their dipole moments.
In particular, the signs of the interactions are reversed and 
their magnitudes are strongly decreased.
The former effect leads to a ``dynamical binding'' of mobile
multiexcitons, and the latter explains the near decoupling 
of excitons in the linear $L=K$ band in the spin-excitation
spectrum at $\nu=1$.

\acknowledgments

AW thanks Manfred Bayer, Leszek Bryja, Pawel Hawrylak, and Marek 
Potemski for helpful discussions and acknowledges support by the 
Polish Ministry of Scientific Research and Information Technology 
under grant 2P03B02424.
This work was also supported by grant DE-FG 02-97ER45657 of the 
Materials Science Program -- Basic Energy Sciences of the U.S. 
Dept.\ of Energy.

\end{document}